\documentclass{article}
\usepackage{graphicx}
\begin{document}
\begin{center}
\textbf{\Large SUB- AND SUPERSONIC HEAT MOTION\\[1ex]
INDUCED BY\\[2ex] FEMTOSECOND LASER PULSES}

\vspace{2cm}
{\large Janina Marciak-Kozlowska}\textit{*}\\
\textit{Institute of Electron Technology}, \textit{Al.} \textit{Lotnik\'{o}w
32-46}, \textit{02-628 Warsaw Poland.}\\

\vspace{1cm}
{\large Miroslaw Kozlowski}\\
\textit{Institute of Experimental Physics and Physics Teacher College},
\textit{Warsaw University}, \textit{Ho\.{z}a 69}, \textit{00-681 Warsaw}, \textit{Poland.}\\
\end{center}

\hbox to 1.5in{\vrule width1.5in height0.4pt depth0in}

\textit{*corresponding author}

\vspace{3cm}
\begin{abstract}
In this paper the superheating of electron plasma by femtosecond
laser pulses is investigated. With Heaviside thermal equation
(\textit{Lasers in Engineering}, \textbf{12,} (2002), p.~17) the
generation of superhot electrons is described. It is shown that in
hot electron plasma (i.e. with electron energies $>5$~MeV)
the thermal shock waves can be generated.
\end{abstract}

\newpage
\section{Introduction}
Recently it has become possible to produce MeV electrons with
short-pulse multiteravat laser system~\cite{1}. The fast ignitor
concept~\cite{2, 3} relevant to the inertial confinement fusion
enhances the interest in this process. In an underdense plasma,
electrons and ions tend to be expelled from the focal spot by the
ponderomotive pressure of an intense laser pulse, and the formed
channel~\cite{4, 5} can act as a propagation guide for the laser
beam. Depending on the quality of the laser beam, the cumulative
effects of ponderomotive and relativistic self focusing~\cite{5}
can significantly increase the laser intensity. For these laser
pulses, the laser electric and magnetic fields reach few hundreds
of GV/m and megagauss, respectively, and quiver velocity in the
laser field is closed to the light speed. The component of the
resulting Lorentz force $(-ev\,x~\,\vec{B})$ accelerates electrons
in the longitudinal direction, and energies of several tens of MeV
can be achieved~\cite{6}. Recently the spectra of hot electrons
(i.e. with energy in MeV region) were investigated. In
paper~\cite{7} the interaction of 500~fs FWHM pulses with CH
target was measured. The electrons with energy up to 20~MeV were
observed. Moreover for electrons with energies higher than 5~MeV
the change of electron temperature was observed: from 1~MeV (for
energy of electrons $<$~5~MeV) to 3~MeV (for energy of electrons
$>5$~MeV). In this paper the interaction of femtosecond laser
pulse with electron plasma will be investigated. Within the
theoretical framework of Heaviside temperature wave equation, the
heating process of the plasma will be described. It will be shown
that in vicinity of energy of 5~MeV the sound velocity in plasma
reaches the value~ $\frac{c}{\sqrt{3} } $ and is independent of
the electron energy.
\section{The model}
The mathematical form of the hyperbolic quantum heat transport was
proposed in~\cite{8} and~\cite{9}. Under the absence of heat or
mass sources the equations can be written as the Heaviside
equations:
\begin{equation}
\frac{1}{v_{\rho }^{2} } \frac{\partial ^{2} \rho }{\partial t^{2}
} +\frac{1}{D_{\rho } } \frac{\partial \rho }{\partial t}
=\frac{\partial ^{2} \rho }{\partial x^{2} }\label{eq1}
\end{equation}
and
    \begin{equation}
    \frac{1}{v_{T}^{2} } \frac{\partial ^{2} T}{\partial t^{2} }
    +\frac{1}{D_{T} } \frac{\partial T}{\partial t} =\frac{\partial
    ^{2} T}{\partial x^{2} }\label{eq2}
    \end{equation}
for mass and thermal energy transport respectively. The discussion
of the properties of Eq.~(\ref{eq1}) was performed in~\cite{8} and
Eq.~(\ref{eq2}) in~\cite{9}. In Eq.~(\ref{eq1}) $v_{\rho } $ ~ is
the velocity of density wave, $D_{\rho } $ is the diffusion
coefficient for mass transfer. In Eq.~(\ref{eq2}) $v_{T} $ is the
velocity for thermal energy propagation and $D_{T} $ is the
thermal diffusion coefficient.

In the subsequent we will discuss the complex transport phenomena,
i.e. diffusion and convection in the external field. The current
density in the case when the diffusion and convection are taken
into account can be written as:
    \begin{equation}
    j=-D_{\rho } \frac{\partial \rho }{\partial t} -\tau
    \frac{\partial j}{\partial t} +\rho V.\label{eq3}
    \end{equation}
In equation~(3) the first term describes the Fourier diffusion,
the second term is the Maxwell-Cattaneo term and the third term
describes the convection with velocity~$V$. The continuity
equation for the transport phenomena has the form:
    \begin{equation}
    \frac{\partial j}{\partial x} +\frac{\partial \rho }{\partial t}
    =0.\label{eq4}
    \end{equation}
Considering both equations~(3) and~(4) one obtains the transport
equation:
    \begin{equation}
    \frac{\partial \rho }{\partial t} =-\tau _{\rho } \frac{\partial^{2} \rho }{\partial t^{2} }
    +D_{\rho } \frac{\partial ^{2} \rho}{\partial x^{2} } -V\frac{\partial \rho }{\partial x}.\label{eq5}
    \end{equation}
In equation~(\ref{eq5}) $\tau _{\rho } $
 denotes the relaxation time for transport phenomena. Let us
perform the Smoluchowski transformation for~ $\rho (x,t)$
    \begin{equation}
    \rho =\exp\left[ \frac{Vx}{2D} -\frac{V^{2} t}{4D} \right] \rho_{1} (x,t).\label{eq6}
    \end{equation}
After substituting~$\rho (x,t)$  formula~(6) to equation~(5) one
obtains for~$\rho _{1} (x,t):$
    \begin{equation}
    \tau _{\rho } \frac{\partial ^{2} \rho _{1} }{\partial t^{2} }
    +\left( 1-\tau _{\rho } \frac{V_{\rho }^{2} }{2D_{\rho } } \right)
    \frac{\partial \rho _{1} }{\partial t} +\tau _{\rho } \frac{V^{4}}{16D_{\rho }^{2} } \rho _{1}
    =D_{\rho } \frac{\partial ^{2} \rho_{1} }{\partial x^{2} }.\label{eq7}
    \end{equation}
Considering that~$D_{\rho } =\tau _{\rho } v_{\rho }^{2} $
equation~(\ref{eq7}) can be written as
    \begin{equation}
    \tau _{\rho } \frac{\partial ^{2} \rho _{1} }{\partial t^{2} }
    +\left( 1-\frac{V_{\rho }^{2} }{2v_{\rho }^{2} } \right)
    \frac{\partial \rho _{1} }{\partial t} +\frac{1}{16\tau _{\rho } }
    \frac{V^{4} }{v_{\rho }^{4} } \rho _{1} =D_{\rho } \frac{\partial^{2} \rho _{1} }{\partial x^{2} }.\label{eq8}
    \end{equation}
In the same manner equation for the temperature field can be
obtained:
    \begin{equation}
    \tau _{T} \frac{\partial ^{2} T_{1} }{\partial t^{2} } +\left(
    1-\frac{V_{T}^{2} }{2v_{T}^{2} } \right) \frac{\partial T_{1}}{\partial t}
     +\frac{1}{16\tau _{T} } \frac{V_{T}^{4} }{v_{T}^{4}} T_{1}
      =D_{T} \frac{\partial ^{2} T_{1} }{\partial x^{2} }.\label{eq9}
    \end{equation}
In equation~(\ref{eq9}) $\tau _{T} ,\;D_{T} ,\,\;V_{T} $
 and
$v_{T} $ are: relaxation time for heat transfer, diffusion
coefficient,
heat convection velocity and thermal wave velocity.

In this paper we will investigate the structure and solution of
the equation~(\ref{eq9}). For the hyperbolic heat transport
Eq.~(\ref{eq9}) we seek a solution of the form:
    \begin{equation}
    T_{1} (x,t)=e^{-\frac{t}{2\tau _{T} } } u(x,t).\label{eq10}
    \end{equation}
After substitution of Eq.~(\ref{eq10}) into Eq.~(\ref{eq9}) one
obtains:
    \begin{eqnarray}
    \tau _{T} \frac{\partial ^{2} u(x,t)}{\partial t^{2} }&-&D_{T}
    \frac{\partial ^{2} u(x,t)}{\partial x^{2} } +\left(
    -\frac{1}{4\tau _{T} } +\frac{V_{T}^{2} }{4D_{T} } +\tau _{T}
    \frac{V_{T}^{4} }{16D_{T}^{2} } \right) u(x,t)\nonumber\\
&-&\tau _{T}
    \frac{V_{T}^{2} }{2D_{T} } \frac{\partial u(x,t)}{\partial t}
    =0\label{eq11}
    \end{eqnarray}
Considering that $D_{T} =\tau _{T} v_{T}^{2} $
 Eq.~(\ref{eq11}) can be written as
    \begin{equation}
    \tau _{T} \frac{\partial ^{2} u}{\partial t^{2} } -\tau _{T} v_{T}^{2}
    \frac{\partial ^{2} u(x,t)}{\partial x^{2} } +\left( -\frac{1}{4\tau _{T}}
     +\frac{V_{T}^{2} }{4\tau _{T} v_{T}^{2} } +\tau _{T} \frac{V_{T}^{4}}
     {16\tau _{T}^{2} v_{T}^{4} } \right) u(x,t)-\frac{V_{T}^{2} }{2v_{T}^{2}}
     \frac{\partial u}{\partial t} =0.\label{eq12}
    \end{equation}
After omitting the term $\frac{V_{T}^{4} }{v_{T}^{4} } $
 in comparison to the term
$\frac{V_{T}^{2} }{v_{T}^{2} } $
 Eq.~(\ref{eq12}) takes the form:
    \begin{equation}
    \frac{\partial ^{2} u}{\partial t^{2} } -v_{T}^{2} \frac{\partial^{2} u}
    {\partial x^{2} } +\frac{1}{4\tau _{T}^{2} } \left(-1+\frac{V_{T}^{2} }{v_{T}^{2} } \right) u(x,t)
    -\frac{V_{T}^{2}}{2v_{T}^{2} \tau _{T} } \frac{\partial u}{\partial t}
    =0.\label{eq13}
    \end{equation}
Considering that $\tau _{T}^{-2} >>\tau _{T}^{-1} $
 one obtains from Eq.~(\ref{eq13})
    \begin{equation}
    \frac{\partial ^{2} u}{\partial t^{2} } -v_{T}^{2} \frac{\partial^{2} u}{\partial x^{2} }
    +\frac{1}{4\tau _{T}^{2} } \left(-1+\frac{V_{T}^{2} }{v_{T}^{2} } \right)
    u=0.\label{eq14}
    \end{equation}
Equation~(\ref{eq14}) is the master equation for heat transfer
induced by ultra-short laser pulses, i.e.~when $\Delta t\approx
\tau _{T} $. In the following we will consider the
Eq.~(\ref{eq14}) in the form:
    \begin{equation}
    \frac{\partial^2 u}{\partial t^2} -v_T^2 \frac{\partial ^2u}{\partial x^2}-qu=0\label{eq15}
    \end{equation}
where
   \begin{equation}
    q=\frac{1}{4\tau^2_T}\left(\frac{V^2_T}{v^2_T}-1\right).\label{eq16}
    \end{equation}
In equation~(\ref{eq16}) the ratio
    \begin{equation}
    M_{T} =\frac{V_{T} }{v_{T} } =\frac{V_{T} }{v_{S}
    }\label{eq17}
    \end{equation}
is the Mach number for thermal processes, for $v_{T} =v_{S} $
 is the sound velocity in the gas of heat carriers~\cite{10}.

\begin{figure}[p]
\centering\includegraphics[height=5.5cm]{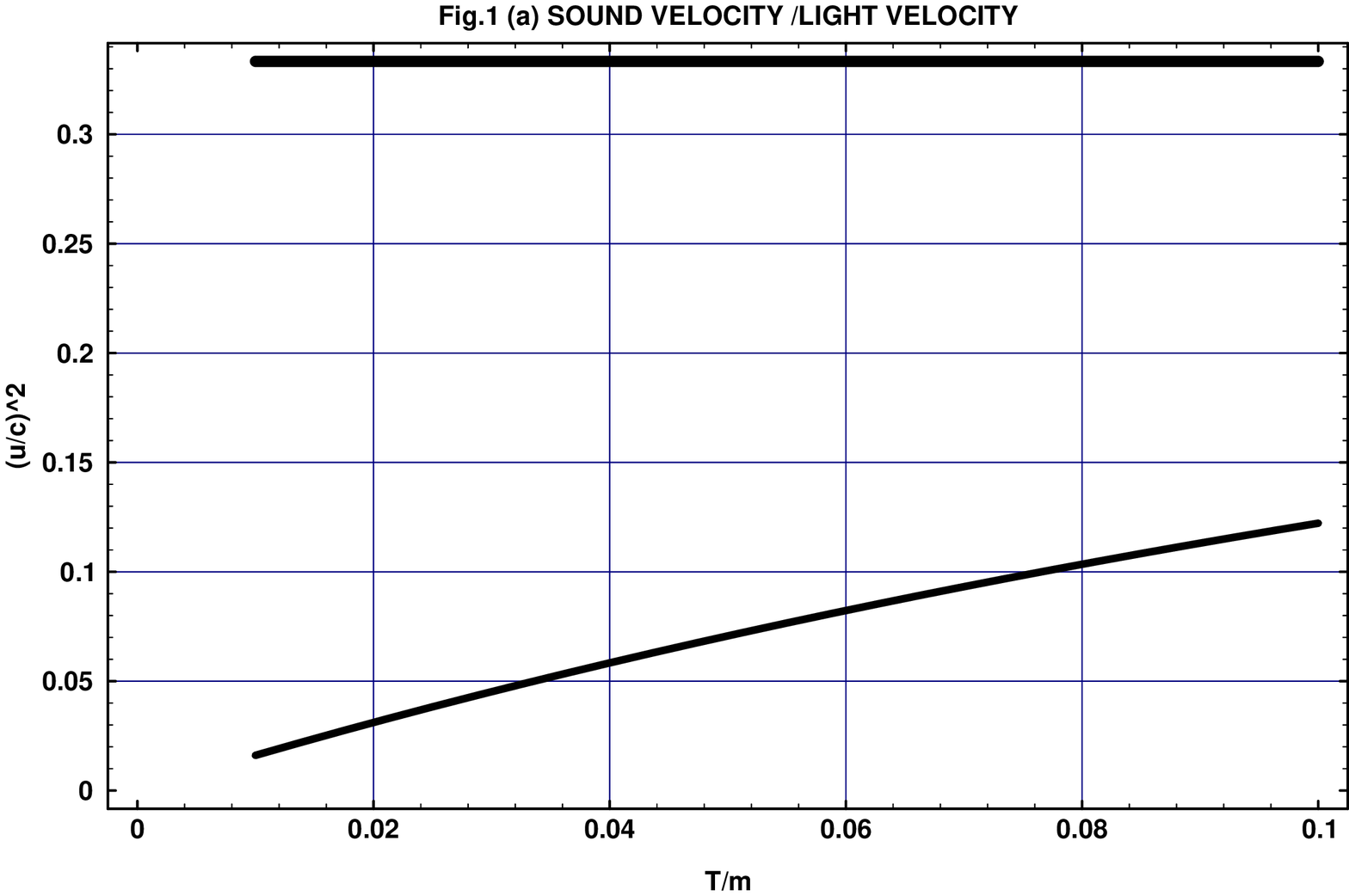}

\centering\includegraphics[height=5.5cm]{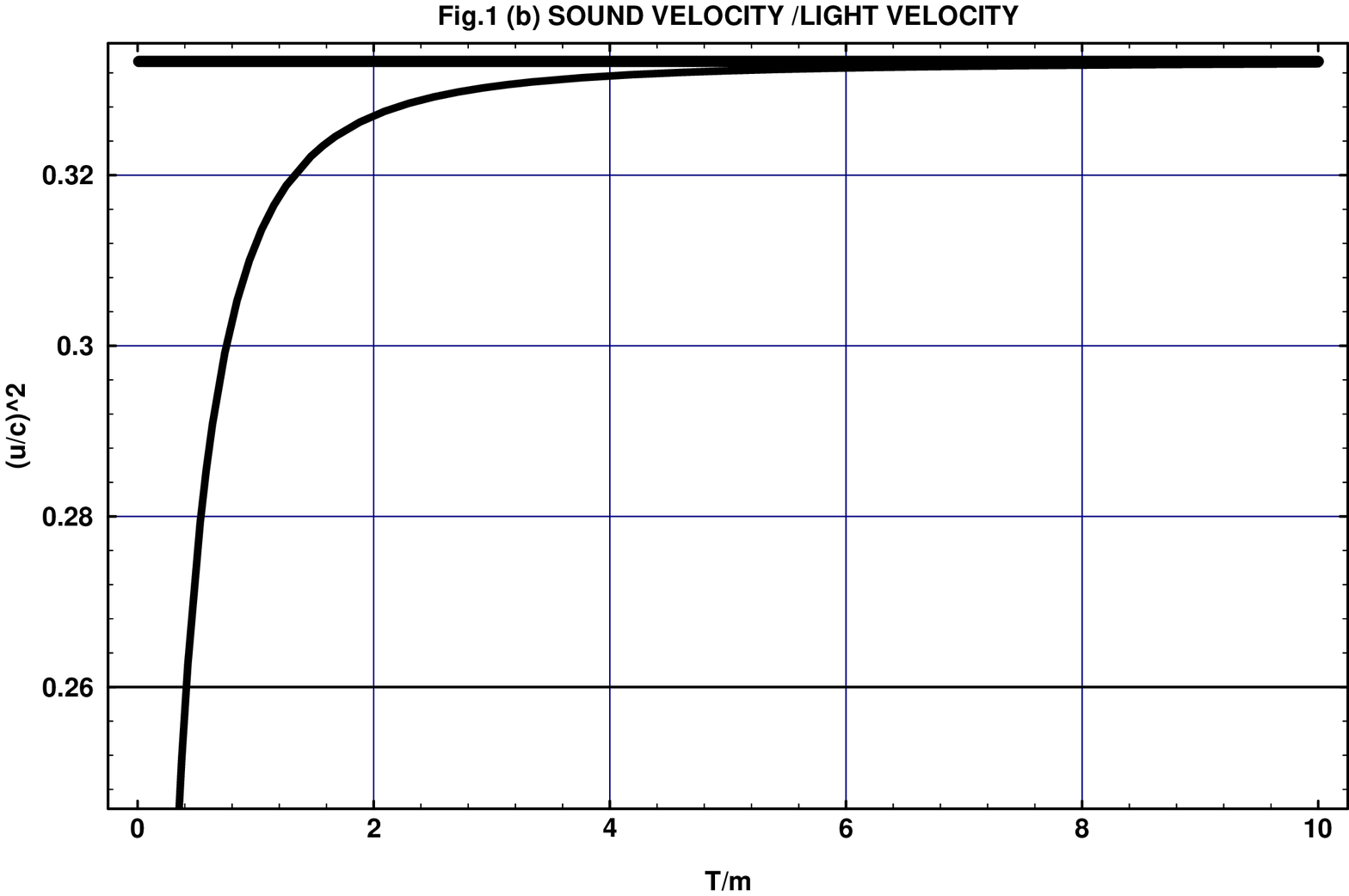}

\centering\includegraphics[height=5.5cm]{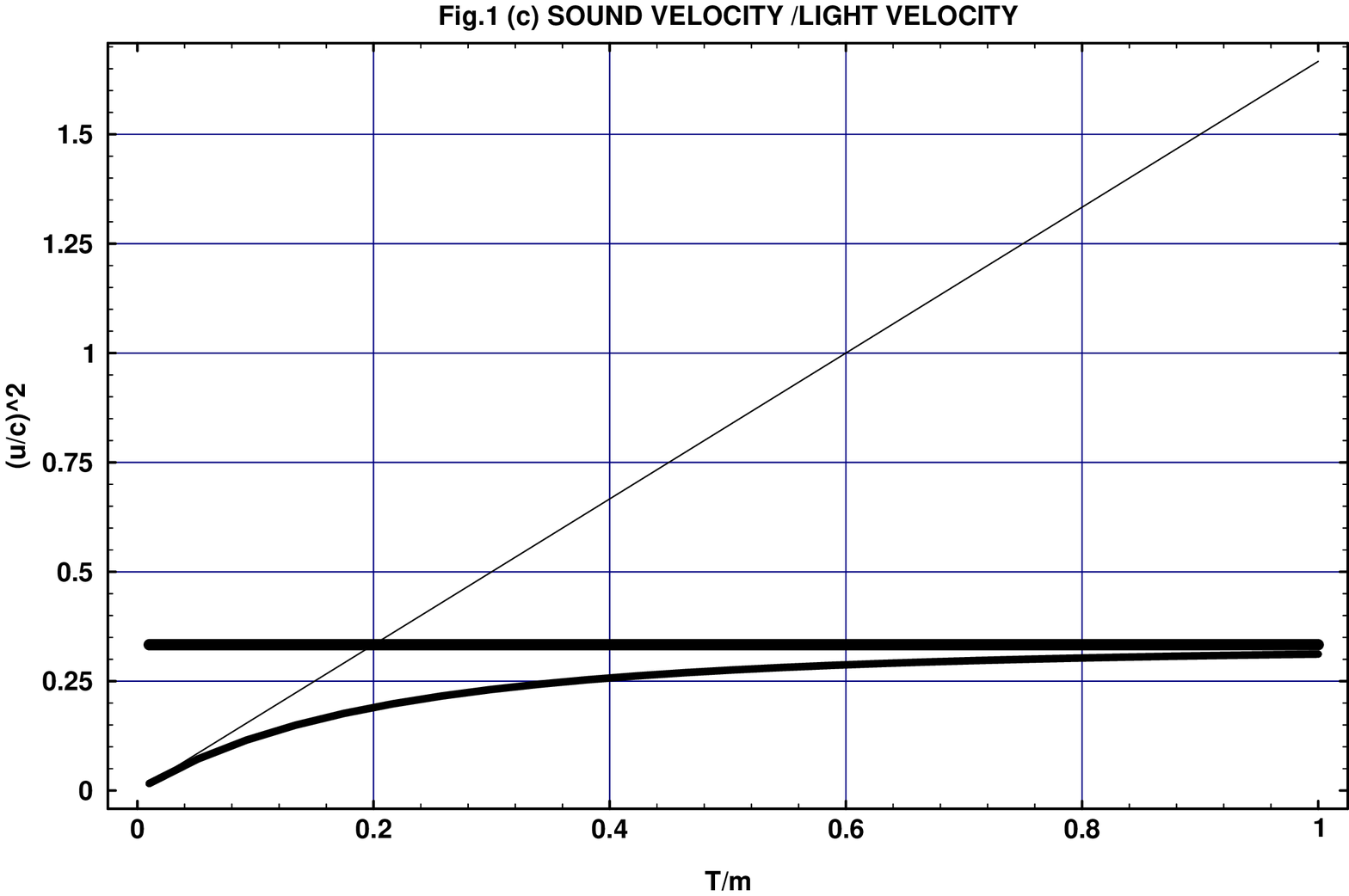} \caption{(a)
The ratio: sound velocity/light velocity $\left( \frac{u}{c}
\right) ^{2} $ as the function $\frac{T}{m} $
 for cold heat carriers
$\left( \frac{T}{m} \ll 1\right) $ ; (b) for hot heat carriers
$\left( \frac{T}{m} \gg 1\right) $ ; (c) Comparison of the ratio
$\left( \frac{u}{c} \right) ^{2} $
 for hot carriers (\vrule width 1em height .8ex depth -.5ex), ultra-relativistic carriers (\vrule width 1em height 1ex depth -.5ex) and Maxwellian
approximation (\vrule width 1em height .6ex depth -.5ex)}\label{fig1}
\end{figure}

\begin{figure}[p]
\centering\includegraphics[height=6cm]{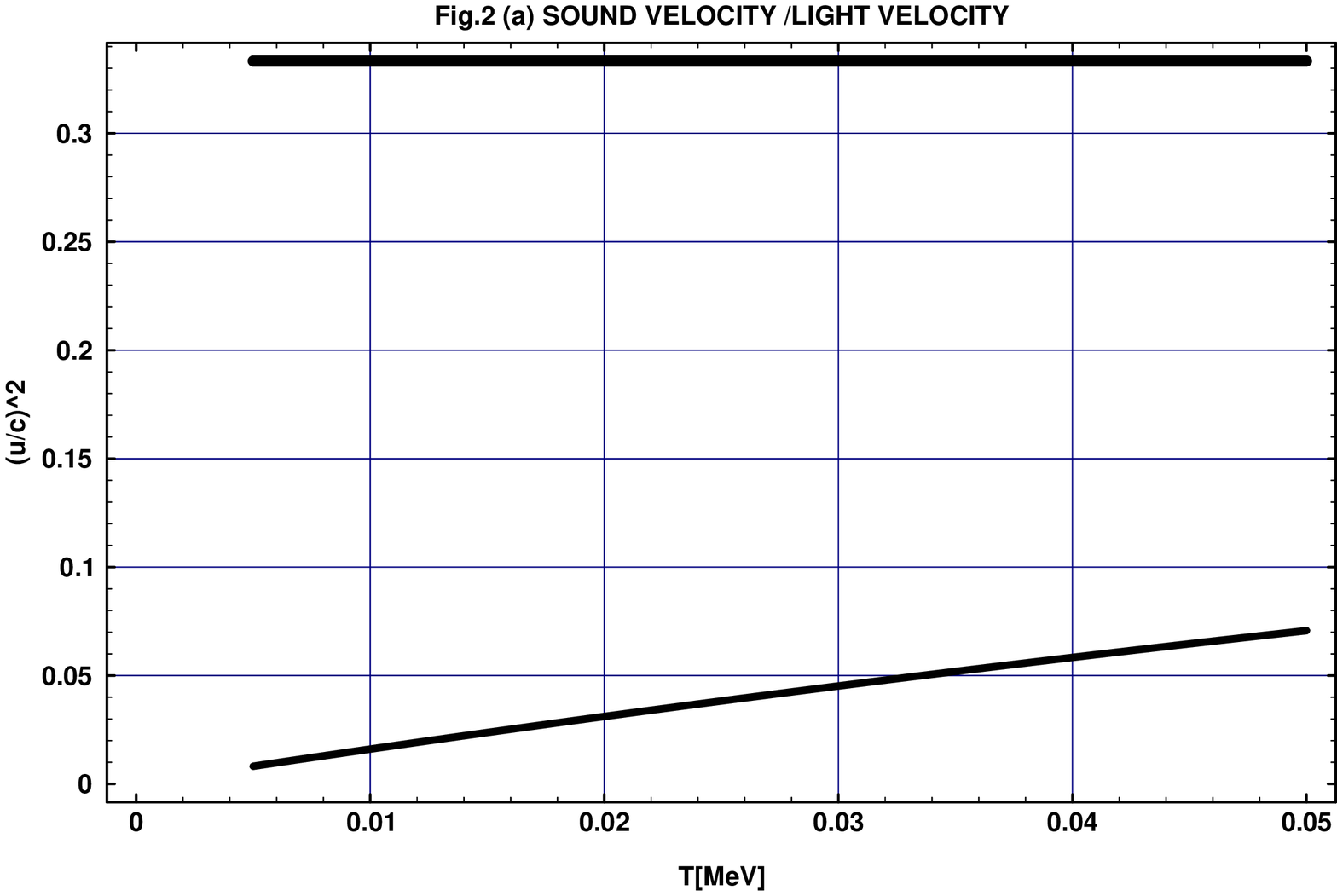}

\centering\includegraphics[height=6cm]{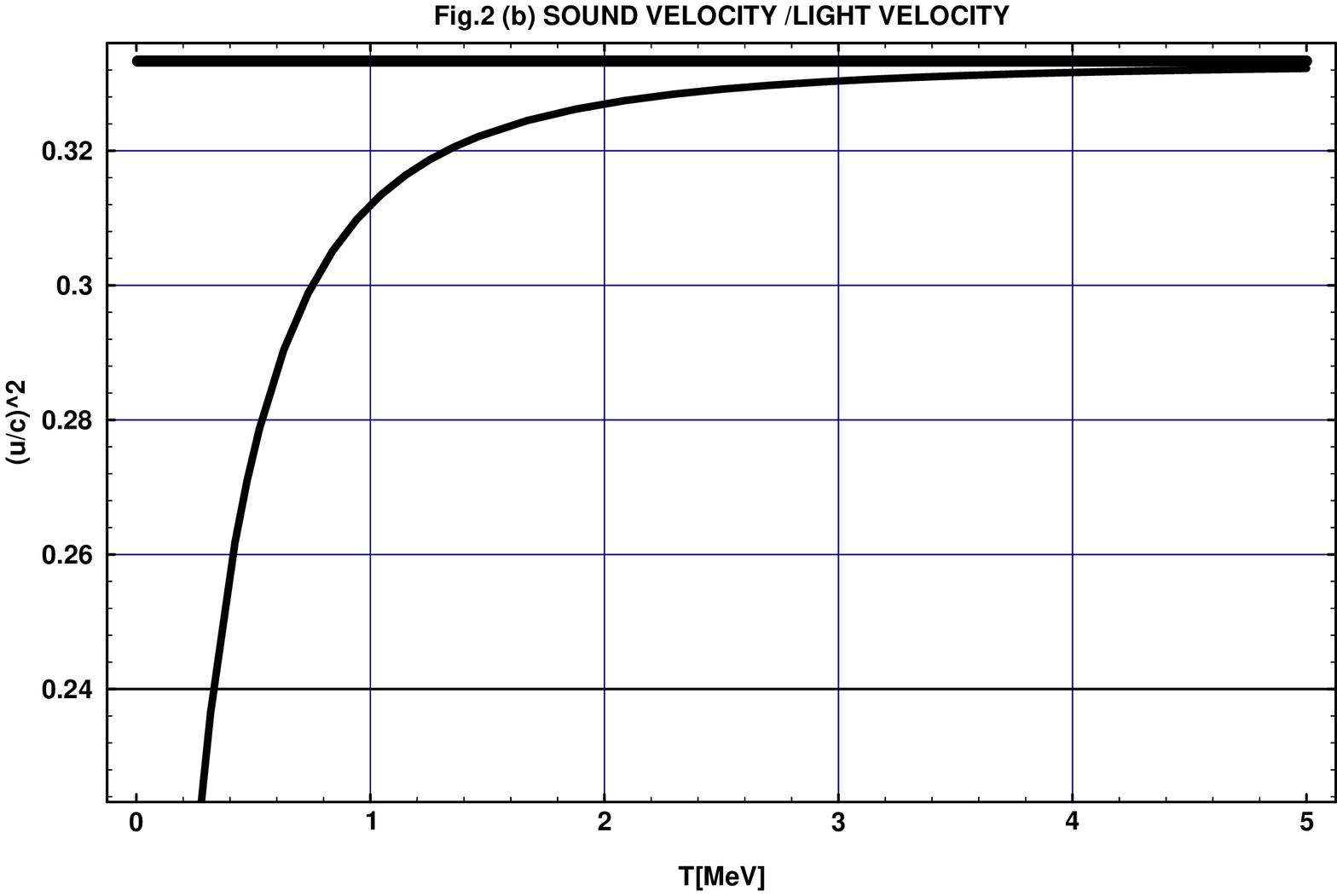}

\centering\includegraphics[height=6cm]{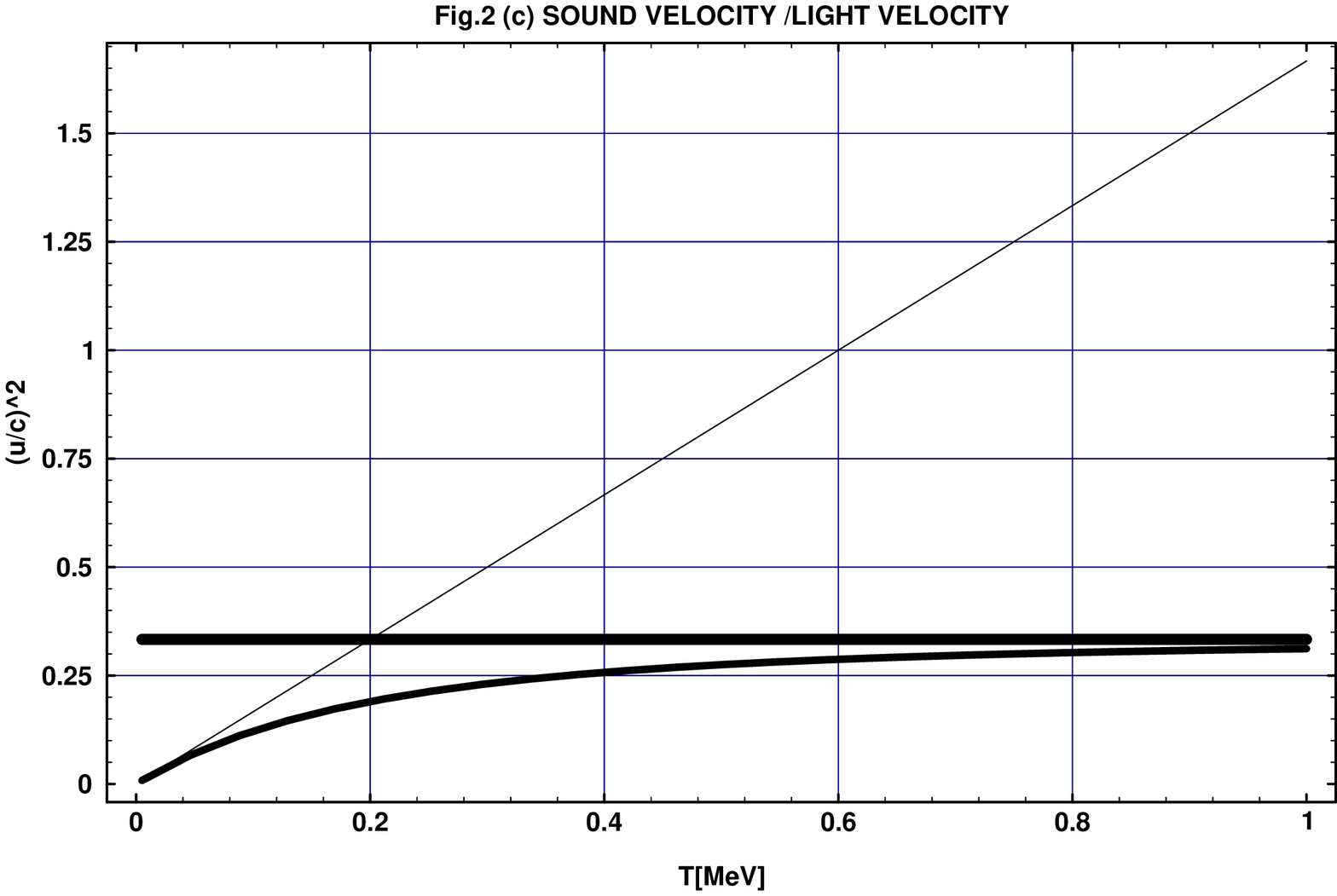} \caption{The
same as in Fig.~\ref{fig1}, but for electrons with mass
m=0.51~MeV/c$^{2}$}\label{fig2}
\end{figure}

\begin{figure}[p]
\centering\includegraphics[height=6cm]{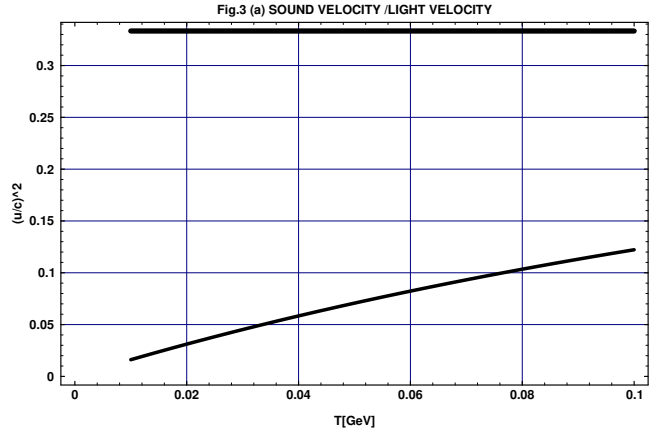}

\centering\includegraphics[height=6cm]{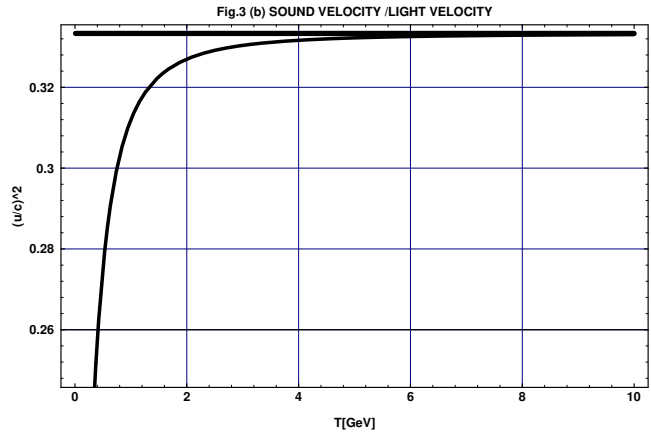}

\centering\includegraphics[height=6cm]{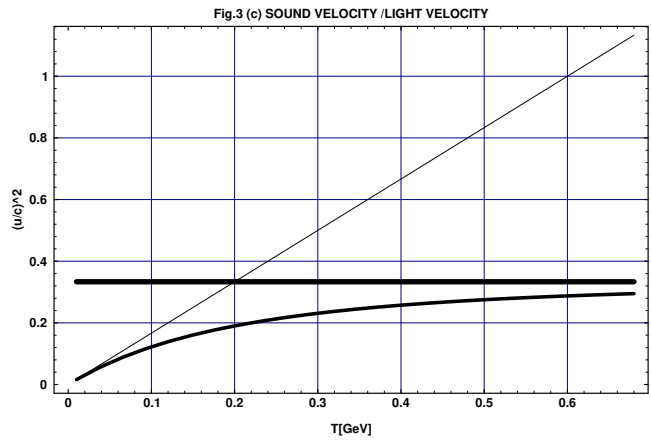} \caption{The
same as in Fig.~\ref{fig2} but for protons with mass
m=0.98~GeV}\label{fig3}
\end{figure}

\begin{figure}[p]
\centering\includegraphics[height=5cm]{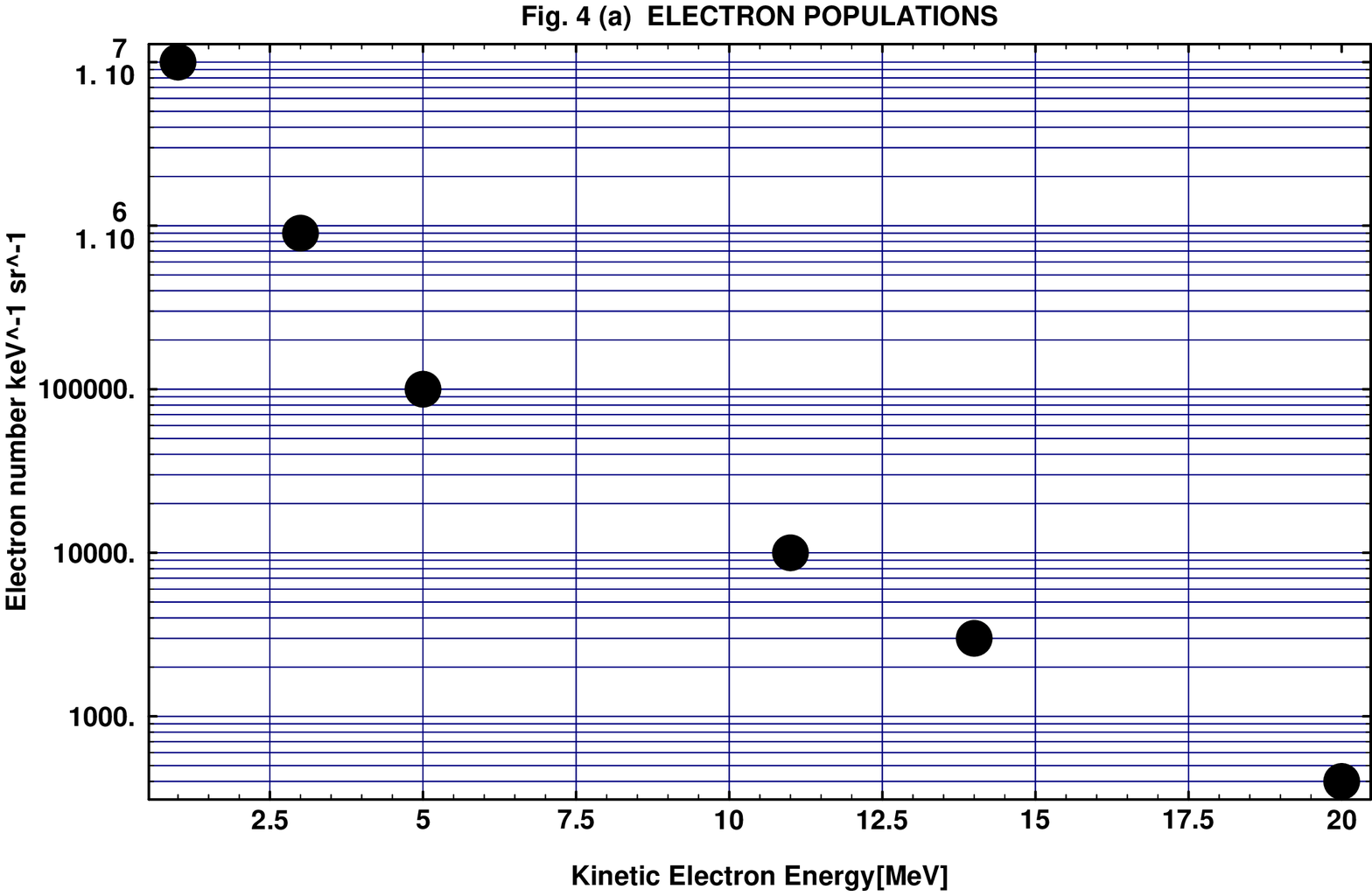}

\centering\includegraphics[height=5cm]{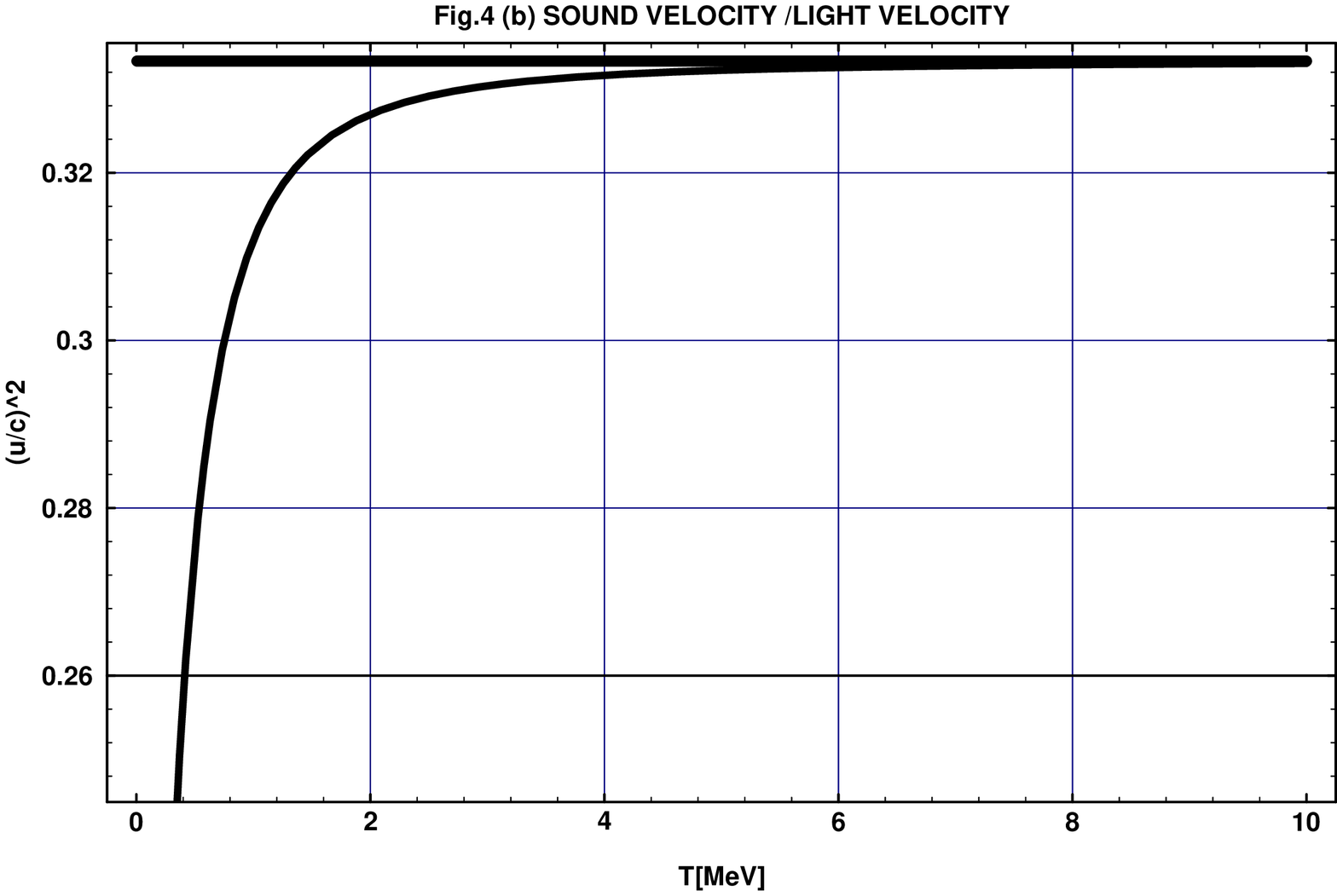}
\caption{(a) The experimental data \cite{7} for the electron
populations. Circles~=~data. The interaction beam intensity is
$I_{\max } \approx 6\cdot 10^{18} \;$W/cm$^{2} $. (b) The ratio~
$\frac{u}{c} $
 as the function of the electron temperature $T$~[MeV]}\label{fig4}
 \end{figure}

In monograph~\cite{10} the structure of equation~(15) was
investigated. It was shown that for $q<0$, i.e. $V_{T} <v_{S} $,
subsonic heat transfer is described by the modified telegrapher
equation
    \begin{equation}
    \frac{1}{v_{T}^{2} } \frac{\partial ^{2} u}{\partial t^{2} }
    -\frac{\partial ^{2} u}{\partial x^{2} } +\frac{1}{4\tau _{T}^{2}
    v_{T}^{2} } \left( \frac{V_{T}^{2} }{v_{S}^{2} } -1\right)
    u=0.\label{eq18}
    \end{equation}
For $q>0,\,\;v_{S} <V_{T} ,$
 i.e. for supersonic case heat transport is described by Klein-Gordon
equation:
    \begin{equation}
    \frac{1}{v_{T}^{2} } \frac{\partial ^{2} u}{\partial t^{2} }
    -\frac{\partial ^{2} u}{\partial x^{2} } +\frac{1}{4\tau _{T}^{2}
    v_{T}^{2} } \left( \frac{V_{T}^{2} }{v_{S}^{2} } -1\right)
    \;u=0.\label{eq19}
    \end{equation}
The velocity of sound~$v_{S} $ depends on the temperature of the
heat carriers. The general formula for sound velocity
reads~\cite{11}:
    \begin{equation}
    v_{S}^{2} =\left( zG-\frac{G}{z} \left( 1+\frac{5G}{z} -G^{2} \right)^{-1} \right)
    ^{-1}.\label{eq20}
    \end{equation}
In formula~(\ref{eq20}) $z=\frac{mc^{2} }{T} $ and~\textit{G} is
of the form~\cite{11}:
    \begin{equation}
    G=\frac{K_{3} (z)}{K_{2} (z)}
    ,\label{eq21}
    \end{equation}
where $c$~is the light velocity, $m$~is the mass of heat carrier,
$T$~is the temperature of the gas and $K_{3} (z),\;K_{2} (z)$ are
modified Bessel functions of the second kind. In Fig.~\ref{fig1} the ratio
of~$\left( \frac{v_{S} }{c} \right) ^{2} $ was presented are the
function of~$\frac{T}{mc^{2} } $. Fig.~\ref{fig1}a presents the $\left(
\frac{u}{c} \right) ^{2} $ for $\frac{T}{mc^{2} } <1$
 (nonrelativistic approximations) and Fig.~\ref{fig1}b presents
the $\left( \frac{u}{c} \right) ^{2}$
 for the very high temperature heat carriers gas, i.e.
$T>mc^{2}$ (relativistic gas). It is interesting to observe that
for nonrelativistic gas, $v_{S}^{2}$ is a linear function of
temperature. From formula~(\ref{eq21}) can be concluded~\cite{11}
that for $T<mc^{2}$ one obtains
    \begin{equation}
    \left( \frac{v_{S} }{c} \right) ^{2} =\left( \frac{5T}{3mc^{2} }
    \right)\label{eq22}
    \end{equation}
i.e as for Maxwellian nonrelativistic gas. On the other hand for
$T\gg mc^{2}$, $\left( \frac{v_{S} }{c} \right) ^{2} =\frac{1}{3}
$
 and is independent of~$T$ where
$v_{S}^{2} =\frac{c^{2} }{3} $ denotes the sound velocity in the
photon gas. In this paper we will study the heat transfer in the
relativistic gas, i.e. when $\left( \frac{v_{S} }{c} \right) ^{2}$
is constant. In that case Eqs.~(\ref{eq18}) and (\ref{eq19}) are
the linear hyperbolic equations. In Fig.~\ref{fig1}(c) it is shown that
Maxwellian approximation is not valid for $\frac{T}{mc^{2} }
>0.05$ and moreover gives a wrong description of
$v_{S} $  for $\frac{T}{mc^{2} } =0.6,$ for $v_{S} >c$ in complete
disagreement with special relativity theory. In Fig.~\ref{fig2}(a,b,c) the
results of calculations of the sound velocity of electron gas and
in Fig.~\ref{fig3}(a,b,c) for proton gas are presented respectively. For
the initial conditions
    $$
    u|_{t=0} =f(x),\quad \left. \frac{\partial u}{\partial t} \right| _{t=0}
    =F(x).
    $$
Solutions of the equation can be find in~\cite{10}:
    \begin{equation}
    u(x,t)=\frac{f(x-v_{T} t)+f(x+v_{T} t)}{2} +\frac{1}{2}
    \int\limits_{x-v_{T} t}^{x+v_{T} t}\Phi (x,t,z)dz,\label{eq23}
    \end{equation}
where
    \begin{eqnarray*}
    \Phi (x,t,z)=\frac{1}{v_{T} } F(z)J_{0} \left( \frac{\sqrt{q} }{v_{T} }
    \sqrt{(z-x)^{2} -v_{T}^{2} t_{2} } \right)\\ +\sqrt{q} tf(z)\frac{J_{0}^{'}
    \left( \frac{\sqrt{q} }{v_{T} } \sqrt{(z-x)^{2} -v_{T}^{2} t^{2} } \right)}
    {\sqrt{(z-x)^{2} -v_{T}^{2} t^{2} } }
    \end{eqnarray*}
and
    $$
    q=\frac{1}{4\tau _{T}^{2} } \left( \frac{V_{T}^{2} }{vT2} -1\right) .
    $$
The general equation for complex heat transfer: diffusion plus
convection can be written as:
    \begin{equation}
    \frac{\partial T}{\partial t} =-\tau _{T} \frac{\partial ^{2} T}{\partial
    t^{2} } +D_{T} \frac{\partial ^{2} T}{\partial x^{2} } -V_{T}
    \frac{\partial T}{\partial x} .\label{eq24}
    \end{equation}
Considering Eqs.~(\ref{eq6}), (\ref{eq10}) and (\ref{eq23}) the
solution of equation~(\ref{eq24}) is
    \begin{eqnarray*}
    T(x,t)&=&\exp\left[ \frac{V_{T} x}{2D} -\frac{V_{T}^{2} t}{4D} \right]
    e^{-\frac{2}{2\tau _{T} } }
    \cdot \bigg( \frac{f(x-v_{T} t)+f(x+v_{T}
    t)}{2}\\ &&\mbox{} +\frac{1}{2} \int\limits_{x-v_{T} t}^{x+v_{T} t}\Phi (x,t,z)dz\bigg).
    \end{eqnarray*}
In Fig.~\ref{fig4} the comparison of the calculation of sound velocity for
electron gas and the electron spectra~[7] is presented. The change
of the shape of the electron spectra in vicinity of 5~MeV can be
easily seen. At the total energy of the 5~MeV the sound velocity
in electron plasma is constant and independent of electron energy.
Electrons with velocities greater than $\frac{1}{\sqrt{3} } c$
 can generate the shock thermal waves which heats the plasma
to higher temperature.

\section{Conclusions}
In paper the Heaviside equation for laser heating electron plasma
was formulated and solved. It was shown that for high energy
electrons, with energy $>5$~MeV the sound velocity in plasma is
constant and equal $v_{S} =\frac{1}{\sqrt{3} } c.$
 The superheating of plasma with electron energy~$>~5$~MeV
can be achieved by the generation of thermal shock waves.

\end{document}